\documentclass[twocolumn,showpacs,preprintnumbers,superscriptaddress,amsmath,amssymb,floatfix, nofootinbib]{revtex4}
\usepackage{graphicx}% Include figure files
\usepackage{dcolumn}% Align table columns on decimal point
\usepackage{bm}% bold math
\usepackage{epstopdf}%eps to pdf figures
\usepackage{color}%use colored text
\usepackage{multirow}
\newcommand{\SU}[1]{\ensuremath{\mathrm{SU}( #1 )}}
\newcommand{\Un}[1]{\ensuremath{\mathrm{U}( #1 )}}
\newcommand{\SO}[1]{\ensuremath{\mathrm{SO}( #1 )}}

\newcommand{\SpR}[1]{\ensuremath{\mathrm{Sp}( #1,\mathbb{R} )}}
\newcommand{\ho}{\ensuremath{\hbar\Omega}}
\begin{document}
%%%%%%%%%%%%%%%%%%%%%%%%%%%%%%%%%%%%%%%%%%%%%%%%%%%%%%%%%%%%%%%%%%%%%%%%%%%%%%%


\title{Efficacy of the SU(3) scheme for {\it ab initio} large-scale calculations \\ beyond the lightest nuclei}

\author{T. Dytrych} 
\email[Corresponding author:]{dytrych@ujf.cas.cz}
\affiliation{Department of Physics and Astronomy, Louisiana State University, Baton Rouge, LA 70803, USA}
\affiliation{Nuclear Physics Institute, Academy of Sciences of the Czech Republic, 250 68 \v{R}e\v{z}, Czech Republic}
\author{P. Maris} 
\affiliation{Department of Physics and Astronomy, Iowa State University, Ames, IA 50011, USA}
\author{K. D. Launey} 
\affiliation{Department of Physics and Astronomy, Louisiana State University, Baton Rouge, LA 70803, USA}
\author{J. P. Draayer} 
\affiliation{Department of Physics and Astronomy, Louisiana State University, Baton Rouge, LA 70803, USA}
\author{J. P. Vary} 
\affiliation{Department of Physics and Astronomy, Iowa State University, Ames, IA 50011, USA}
\author{D. Langr}
\affiliation{Faculty of Information Technology, Czech Technical University, Prague 16000, Czech Republic}
\affiliation{Aerospace Research and Test Establishment, Prague 19905, Czech Republic}
\author{E. Saule}
\affiliation{Department of Computer Science, University of North Carolina at Charlotte, NC 28223, USA}
\author{M. A. Caprio}
\affiliation{Department of Physics, University of Notre Dame, Notre Dame, IN 46556, USA}
\author{U. Catalyurek}
\affiliation{Department of Biomedical Informatics, The Ohio State University, Columbus, OH 43210, USA}
\affiliation{Department of Electrical and Computer Engineering, The Ohio State University, Columbus, OH 43210, USA}
\author{M. Sosonkina}
\affiliation{Department of Modeling, Simulation and Visualization Engineering, Old Dominion University, Norfolk, VA 23529, USA}

\date{\today}

%%%%%%%%%%%%%%%%%%%%%%%%%%%%%%%%%%%%%%%%%%%%%%%%%%%%%%%%%%%%%%%%%%%%%%%%%%%%%%%
\begin{abstract}
%%%%%%%%%%%%%%%%%%%%%%%%%%%%%%%%%%%%%%%%%%%%%%%%%%%%%%%%%%%%%%%%%%%%%%%%%%%%%%%
We report on the computational characteristics of {\it ab initio} nuclear
structure calculations in a symmetry-adapted no-core shell model (SA-NCSM)
framework.  We examine the computational complexity of the current
implementation of the SA-NCSM approach, dubbed {\tt LSU3shell}, by analyzing
{\it ab initio}  results for $^{6}$Li  and $^{12}$C in large harmonic
oscillator model spaces and \SU{3}-selected subspaces.  We demonstrate
{\tt LSU3shell}'s strong-scaling properties  achieved with highly-parallel
methods for computing the many-body matrix elements.  Results compare favorably
with complete model space calculations and significant memory savings are
achieved in physically important applications. In particular, a well-chosen
symmetry-adapted basis affords memory savings in calculations of states with a
fixed total angular momentum in large model spaces while exactly preserving translational
invariance.
\end{abstract}
\maketitle

%%%%%%%%%%%%%%%%%%%%%%%%%%%%%%%%%%%%%%%%%%%%%%%%%%%%%%%%%%%%%%%%%%%%%%%%%%%%%%%
\section{Introduction and Motivation}
%%%%%%%%%%%%%%%%%%%%%%%%%%%%%%%%%%%%%%%%%%%%%%%%%%%%%%%%%%%%%%%%%%%%%%%%%%%%%%%

% Some paragraphs are nearly identical with those in the MCSM paper

%A long-standing goal of nuclear physics is to obtain the exact solutions of a
%realistic Hamiltonian for finite nuclei, a Hamiltonian that describes well the
%few-body data, and to compare those results with experiment where available.
%Once validated, the resulting methods will be very useful for predicting
%properties of nuclei that cannot be studied experimentally but may be of great
%importance in understanding astrophysical phenomena or for practical
%applications such as energy generation.   This is the physics program we aim
%to empower by developing and testing new many-body methods.

In the last few years, {\it ab initio} approaches to nuclear structure and
reactions have considerably advanced our understanding and capability of
achieving  first-principles descriptions of $p$-shell nuclei
(see Refs. \cite{GFMC0, GFMC1, GFMC2,NCSM12a,NCSM12b,NCSM12c,CC,RothN07,EpelbaumKLM11} and references therein).  At the same time, fundamental approaches to the
nucleon-nucleon ($NN$) and three-nucleon ($NNN$) interactions, such as
meson-exchange theory and chiral effective field theory,  have yielded major
progress \cite{Wiringa:1994wb,Pieper_3NF,Illinois,Epelbaum,N3LO, N3LOb}.  Successful
realistic $NN$ interactions from inverse scattering have also emerged
\cite{Shirokov07a,Shirokov07b}. These new developments in microscopic nuclear theory combine
to place serious demands  on available computational resources for achieving
converged properties of $p$-shell nuclei.  This points to the need of further
major advances  in many-body methods to access a wider range of nuclei and
experimental observables, while retaining the {\it ab initio} predictive power.

% Paragraph on the motivation for a symmetry-adapted basis

These considerations motivate us to develop and investigate a novel model, the
{\it ab initio} symmetry-adapted no-core shell model (SA-NCSM)
\cite{DytrychLMCDVL_PRL13}, which, by taking advantage of symmetries inherent to
the nuclear dynamics \cite{Elliott, Elliott2,Sp3R, RoweRPP}, can provide access to  heavier nuclei
and larger model spaces essential to accommodate collective, deformed, and
cluster substructures
\cite{Harvey68,RosensteelR80,DraayerW83,DraayerWR84,RoweRPP,Suzuki86_89a,Suzuki86_89b, DreyfussLDDB13, TobinFLDDDB14, LauneyDDDB14,LauneyDDSD15}. This
is achieved by recognizing that the choice of a basis is crucial  and that the
SA-NCSM affords a solution that is a linear combination of a limited number of
basis states of definite nuclear deformation.  This yields memory savings in
larger model spaces with fixed total angular momentum.

The concept underpinning the SA-NCSM has been demonstrated in our recent {\it
ab initio} studies of properties of $^{6}$Li, $^{6}$He, and $^{8}$Be
\cite{DytrychLMCDVL_PRL13, DytrychHLDMVLO15}.  The potential gains from the
SA-NCSM have also been demonstrated in our earlier study where we found that
{\it ab initio} wavefunctions of $^{12}$C and $^{16}$O calculated by the
no-core shell model (NCSM)  \cite{NCSM12a} project  well onto a symmetry-adapted
subspace that is only a tiny fraction of the corresponding complete model space
\cite{SA-NCSM1,DytrychSBDV_PRCa07,SA-NCSM2}.  While the SA-NCSM states can be obtained through a
unitary transformation from the basis used in the NCSM, and hence span the
entire space,  the growth of the model space within the SA-NCSM framework can
be managed, as shown here, by down-selecting to the physically relevant states
as determined through symmetry considerations. This brings forward the major
distinction as compared to the NCSM, namely, the selection of the SA-NCSM
many-nucleon basis states retained in the calculation, which, in turn, allows the
SA-NCSM to accommodate heavier nuclei and larger model spaces.  Further
advantages of the SA-NCSM include the capability of exactly removing the spurious
center-of-mass motion within a selected subspace (thereby retaining this
important factorization feature of the NCSM), as well as the use of efficient
group-theoretical methods and coupling rules to build the many-nucleon basis
states and interaction matrix elements.

In the present work, 
we study the efficacy of the SA-NCSM basis by comparing the complexity of
calculations for SA-NCSM and NCSM, and we investigate the significance of
selected SA-NCSM spaces against the corresponding NCSM (complete-space SA-NCSM)
results.  To achieve this, we 
evaluate properties of $^{12}$C with  both methods for
% compare results with No Core Full Configuration (NCFC) method \cite{Maris09_NCFC} with
comparable cutoffs.  
 We adopt the JISP16 $NN$ interaction \cite{Shirokov07a,Shirokov07b} without
renormalization and we neglect $NNN$ interactions.  For both many-body methods,
all $A$ nucleons in the nucleus are treated on the same footing. 
Observables, that can be experimentally measured, are obtained from $A$-nucleon wavefunctions resulting from
Hamiltonian diagonalization in the chosen many-body model space.  
%To perform the comparisons between the methods, we focus on selected properties of $^{12}$C. 
In particular, we consider the binding energies of the ground state ($gs$) and eight low-lying excited states of positive parity in $^{12}$C. 
%three lowest excited states of each spin-parity  -- $0^+$ and $2^+$. (DO WE WANT TO ADD THE LOWEST 4+ TO ROUND OUT THE GS ROTATIONAL BAND?). 
We also compare the $gs$ point-particle matter  root-mean-square (rms)
%, charge and neutron 
radius as well as the mass quadrupole moment of the first excited $2^+$ and $4^+$ states and the  $B(E2)$ electric quadrupole transition strength from the $2^+_1$
%matrix element for the electric quadrupole transition 
to the ground state, together with the $B(M1; 1_1^+ \rightarrow 0^+_{\rm gs})$ transition strength and the $1_1^+$ magnetic dipole moment.

It is worth noting that there are additional efforts aimed at accelerating the
convergence of {\it ab initio} no-core many-body methods using basis function
techniques. The ``Importance-Truncated" no-core shell model (IT-NCSM)
\cite{RothN07} attempts to sample the many-body configurations above a given
cutoff using perturbative estimates of their contributions to the energy of
low-lying states.  The {\it ab initio} no-core Monte Carlo Shell Model
(NC-MCSM) \cite{Otuska_MCSM1,Otuska_MCSM2,Otuska_MCSM3} aims to sample model spaces generated from an
angular momentum and parity projected Hartree-Fock basis.  Alternatively, the
no-core full configuration approach  (NCFC) represents an extrapolation to the
infinite matrix limit of a sequence of calculations in finite model spaces  and
provides important observables, as some calculated quantities  monotonically
approach the exact result with increasing model spaces \cite{Maris09_NCFCa, Maris09_NCFCb, Maris09_NCFCc, Maris09_NCFCd}.  In
this connection, there has been substantial recent progress in developing
improved extrapolation techniques with quantitative uncertainty estimates
\cite{Coon:2012ab,Furnstahl:2012qg,KonigBFMP14,WendtFPS15,OdellPP15}. 

%It remains to be seen which method, among the many under investigation, will be more efficient and for which systems and which observables.    Outstanding challenges include the fully microscopic description of clustering phenomena and extensions to {\it ab initio} nuclear reaction theory.

%Following moved here from below:
Our goal is, first, to study the accuracy of SA-NCSM model space selections as
compared to the  complete-space results, as well as the dependence of the
results on the model space parameters for $^{12}$C (Sec.
\ref{sec:C12properties}), and, then, to investigate the efficacy of the SA-NCSM
by comparing the complexity of NCSM  and SA-NCSM calculations (Sec.
\ref{sec:complexity}).  This, in turn, will reveal the significance, in terms
of potential memory savings for fixed-$J$ bases, of reducing the complete model
space by searching for and retaining the physically relevant basis
states important for the low-energy properties of nuclei with a realistic $NN$
interaction.

%%%%%%%%%%%%%%%%%%%%%%%%%%%%%%%%%%%%%%%%%%%%%%%%%%%%%%%%%%%%%%%%%%%%%%%%%%%%%%%
\section{Quantum Many-Body Methods Adopted}
%%%%%%%%%%%%%%%%%%%%%%%%%%%%%%%%%%%%%%%%%%%%%%%%%%%%%%%%%%%%%%%%%%%%%%%%%%%%%%%

For a general problem, both NCSM and SA-NCSM adopt the intrinsic
non-relativistic nuclear plus Coulomb interaction Hamiltonian defined as
follows: \begin{equation} {H = T_{\rm rel} + V_{NN}  + V_{NNN} + \ldots +
V_{\rm Coulomb}, \label{intH}} \end{equation} where the $V_{NN}$
nucleon-nucleon and $V_{NNN}$ 3-nucleon interactions are included along
with the Coulomb interaction between the protons.  The Hamiltonian may
include additional terms such as multi-nucleon interactions among more than
three nucleons simultaneously and higher-order electromagnetic interactions
such as magnetic dipole-dipole terms.  

In this study, the JISP16 $NN$ interaction is adopted. It produces a
high-quality description of the $NN$ scattering data and the deuteron
\cite{Shirokov07a,Shirokov07b} as well as a good description of a range of properties of
light nuclei \cite{Maris09_NCFCa}.  We include the Coulomb interaction but we
neglect  higher-order electromagnetic interactions as well as $NNN$
interactions  and beyond. We treat neutron and proton orbitals independently so
total isospin is not conserved.

For each method, we retain many-nucleon basis states of a fixed parity, consistent
with the Pauli principle, and limited by a many-body basis cutoff $N_{\rm
max}$. The $N_{\rm max}$ cutoff is defined as the maximum number of harmonic
oscillator (HO) quanta allowed in a many-nucleon basis state above the minimum for
a given nucleus.  We seek to obtain the lowest few eigenvalues and
eigenfunctions of the Hamiltonian (\ref{intH}).  The resulting calculated $gs$
energy is a rigorous upper bound on the exact result for the full (infinite)
basis and monotonically approaches the exact result as the $N_{\rm max}$
increases. This upper bound character applies to the lowest calculated state of
each total angular momentum and parity.  We defer extrapolations to exact
infinite matrix results to other efforts such as Refs.
\cite{Maris09_NCFCa, Maris09_NCFCb,Maris09_NCFCc, Maris09_NCFCd,Coon:2012ab,Furnstahl:2012qg}.

The NCSM calculations may be performed in an $M$-scheme basis where the
many-nucleon basis states are constructed with a good total magnetic projection
$M$ that is the same for all basis states ($M=0$ here).  The eigensolutions
 have good total angular momentum up to numerical errors and this serves as
a cross-check on the precision of the calculations. Note that we employ a
Lanczos scheme that selectively converges the low-lying solutions, regardless
of their total angular momenta, in the same $M$-scheme calculation, thus facilitating the identification of the ground-state spin as well as calculations of electromagnetic transition rates.
For the NCSM $M$-scheme calculations reported here, we employ the code
``Many-Fermion Dynamics - nuclear" or ``{\tt MFDn}" \cite{Vary92_MFDn} which
has been revised and optimized for leadership-class parallel computers
\cite{MFDn1, MFDn2, MFDn3, MFDn4}.  

The NCSM calculations may also be performed in a  $J$-scheme basis where the
many-nucleon basis states are constructed with a good total angular momentum $J$
and total magnetic projection $M$ that is the same for all basis states
\cite{Aktulga}. This approach again uses the Lanczos procedure.  Several runs
are then needed to map out all the states in the low-lying spectrum. As the
SA-NCSM utilizes basis states with a good $J$ quantum number, we use the {\it
J}-scheme NCSM approach for the primary reference case in order to compare
computational complexity. It is useful to note, for example, that one may need
a fixed-$J$ scheme to isolate members of excited rotational bands that reside in
a dense spectrum since the Lanczos procedure in the $M$ scheme is less efficient
for these states.

For the SA-NCSM calculations, we work in a many-nucleon basis labeled by
the SU(3) quantum numbers $(\lambda\,\mu)$ of the Elliott model \cite{Elliott, Elliott2},
orbital momentum $L$, proton, neutron, and total intrinsic spins ($S_{p}$,
$S_{n}$, and $S$), as well as  total angular momentum $J$ and its projection
$M$ (similarly to NCSM, $M=0$ here). We employ a new code, ``{\tt LSU3shell}"
\cite{LSU3shell15}, that runs efficiently on parallel computers, as shown
in this paper.  
%SA-NCSM calculations in a complete  model space for a given $N_{\max}$  yield results that coincide with those obtained by the NCSM in the same $N_{\max}$ space.

%%%%%%%%%%%%%%%%%%%%%%%%%%%%%%%%%%%%%%%%%%%%%%%%%%%%%%%%%%%%%%%%%%%%%%%%%%%%%%%
\subsection{Model Space Parameters: $N_{\rm max}$ and $\hbar\Omega$}
%%%%%%%%%%%%%%%%%%%%%%%%%%%%%%%%%%%%%%%%%%%%%%%%%%%%%%%%%%%%%%%%%%%%%%%%%%%%%%%

All solutions have a dependence on the cutoff $N_{\rm max}$ and on the HO
energy $\hbar\Omega$.  The degree to which we obtain results independent of the
$N_{\rm max}$ and of $\hbar\Omega$ is a measure of the convergence of the
results since fully converged results are independent of both basis parameters.  

For both methods, we employ the many-body $N_{\rm max}$ truncation where we
enumerate all many-body states, with the selected symmetries, possessing total
HO excitation quanta less than or equal to $N_{\rm max}$.  Specifically, each
single-particle state in a many-nucleon basis state contributes $2n+l$ to the
total HO quanta ($n$ is the radial quantum number and $l$ is the orbital
angular momentum quantum number) for that basis state. Then, the minimum sum
for a given nucleus (of the lowest allowed configuration) is subtracted to give
the total HO  excitation  quanta.  The smallest model space for each nucleus is
then $N_{\rm max}=0$ and increases in units of $2$ for the states of the same
parity.  Odd values of $N_{\rm max}$ cover the states with opposite parity.
With this scheme, basis states where one nucleon carries all the $N_{\max}$
quanta are included, in which cases one nucleon occupies the highest HO shell.

In both NCSM and SA-NCSM, the $N_{\rm max}$ cutoff in the HO basis is valuable
for preserving Galilean invariance.  That is, with the $N_{\rm max}$ cutoff, we guarantee that all solutions factorize into a product of
intrinsic and center-of-mass motion components.  With a Lagrange multiplier term acting on the center-of-mass
\cite{Lipkin58,Verhaar60,Lawson74,NCSM12a}, we remove states of center-of-mass excitation from low-lying spectrum.
All observables may then be evaluated free of spurious center-of-mass motion contributions.

%%%%%%%%%%%%%%%%%%%%%%%%%%%%%%%%%%%%%%%%%%%%%%%%%%%%%%%%%%%%%%%%%%%%%%%%%%%%%%%
\subsection{{\it Ab initio} SA-NCSM}
%%%%%%%%%%%%%%%%%%%%%%%%%%%%%%%%%%%%%%%%%%%%%%%%%%%%%%%%%%%%%%%%%%%%%%%%%%%%%%%
The SA-NCSM framework allows one to down-select from all possible
configurations to a subset that tracks with an inherent preference of a system
towards low-spin and high-deformation dominance -- and symplectic multiples
thereof in high-$N_{\max}$~spaces \cite{DytrychLMCDVL_PRL13} -- as revealed to be important in
realistic NCSM wavefunctions \cite{SA-NCSM1, DytrychSBDV_PRCa07,SA-NCSM2}. 

The  many-nucleon basis states of the SA-NCSM are decomposed into spatial and
intrinsic spin parts, where the spatial part is further classified according to the
\SU{3}$\supset$\SO{3} group chain.  The significance of the \SU{3} group for a
microscopic description of the nuclear collective dynamics can be seen from the
fact that it is the symmetry group of the successful Elliott
model~\cite{Elliott, Elliott2}, and a subgroup of the physically relevant \SpR{3}
symplectic model~\cite{Sp3R, RoweRPP}, which provides a comprehensive theoretical
foundation for understanding the dominant symmetries of nuclear collective
motion.
The SA-NCSM basis states are labeled schematically as
\begin{equation} 
|\vec{\gamma}; N(\lambda\,\mu)\kappa L; (S_{p}S_{n})S; J M\rangle,  \label{SAbasis} 
\end{equation}
where $S_{p}$, $S_{n}$, and $S$ denote proton, neutron, and total intrinsic
spins, respectively, $N$ is the total number of HO excitation quanta, and
$(\lambda\,\mu)$ represent a set of quantum numbers that labels an \SU{3}
irreducible representation, or 
``irrep''\footnote{The \SU{3} irrep labels $(\lambda\,\mu)$ bring forward important information about nuclear shapes and
deformation, according to an established mapping
\cite{CastanosDL88,RosensteelR77,LeschberD87}. For example, $(0 0)$,
$(\lambda\, 0)$ and $(0\,\mu)$ describe spherical, prolate and oblate
deformation, respectively.
}. 
The label $\kappa$ distinguishes
multiple occurrences of the same orbital momentum $L$ in the parent irrep
$(\lambda\,\mu)$.  The $L$ is coupled with $S$ to the total angular momentum
$J$ and its projection $M$.    The symbol $\vec{\gamma}$ schematically denotes the
additional quantum numbers needed to specify a distribution of nucleons over
the major HO shells and their single-shell and  inter-shell quantum numbers.
Specifically, in each major HO shell $\eta$ with degeneracy $\Omega_{\eta}$,
protons (or neutrons)  are arranged into antisymmetric
$\Un{\Omega_{\eta}}\times \SU{2}_{S_{\eta}}$ irreps~\cite{DraayerLPL89}, with
$\Un{\Omega_{\eta}}$ further reduced with respect to SU(3), providing the single-shell labels
$\left[f_{1},\dots, f_{\Omega_{\eta}}\right] \alpha_{\eta}
(\lambda_{\eta}\,\mu_{\eta}) S_{\eta} $. Note that a
spatial symmetry
% the set of $ (\lambda_{\eta}\,\mu_{\eta})$ configurations,
associated with a Young tableau $\left[f_{1},\dots, f_{\Omega_{\eta}}\right]$
is uniquely determined by the imposed antisymmetrization and the associated
intrinsic spin $S_{\eta}$~\cite{DraayerLPL89}.  A multiplicity index $\alpha_{\eta}$ is required to
distinguish multiple occurrences of \SU{3} irrep $(\lambda_{\eta}\,\mu_{\eta})$
in a given $\Un{\Omega_{\eta}}$ irrep. Coupling of these single-shell
configurations further yield inter-shell \SU{3}$\times\SU{2}_{S}$ quantum
numbers for protons and for neutrons; the proton and neutron  configurations
are finally coupled to good $(\lambda\,\mu)\kappa LS;JM$.  All of these labels
uniquely determine the SA-NCSM basis states~(\ref{SAbasis}).  
%------------------------------------------------------------------------------
% Fig: model space selection 
%------------------------------------------------------------------------------
\begin{figure*}[th]
\begin{center}
\includegraphics[width=0.85\textwidth]{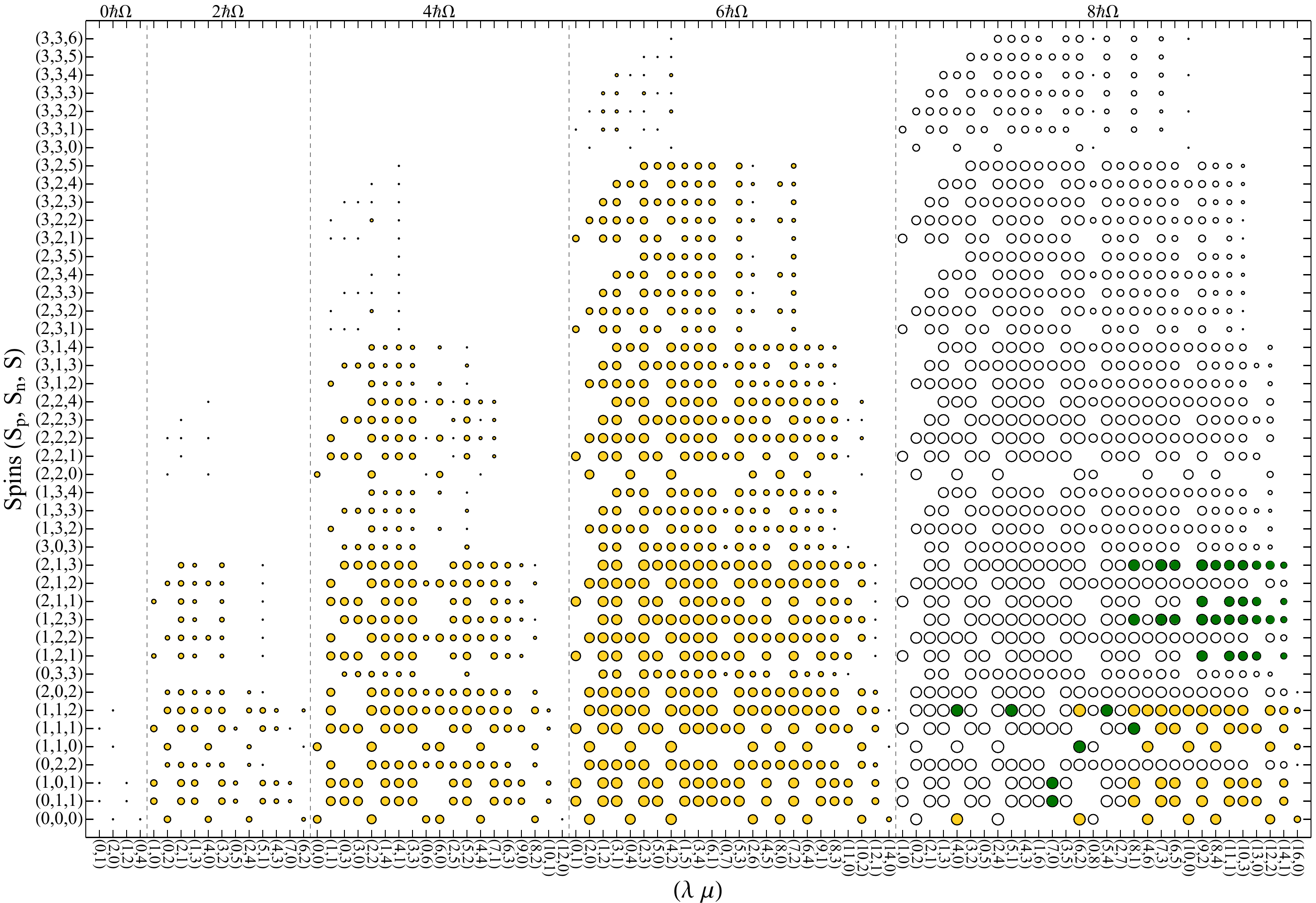}
\caption{ 
Complete $J=0$ model space for $^{12}$C and $N_{\rm max}=8$ given in terms of proton ($S_p$), neutron ($S_n$) and total ($S$) spin values (vertical axis) and deformation $(\lambda\,\mu)$ (horizontal axis). Each circle represents basis states carrying the same $(S_p, S_n, S)$ and $(\lambda\,\mu)$ quantum numbers within a \ho-subspace, with the radius being proportional to log$_{10}$ of the number of such states. Filled light yellow circles indicate the $\langle 6 \rangle8$-A restricted model space employed by the SA-NCSM for calculations of $0^+$ states in $^{12}$C, while all  filled circles indicate the $\langle 6 \rangle8$-B restricted model space.  Similar selection rules are applied to higher-$J$ states. 
}
\label{Fig:12C_ space_selection}
\end{center}
\end{figure*}
%%Tomas, I had to redo the pdf file you sent me (for reviewing purposes), as the green circles did not show up in the latex-generated pdf. But if you could use a better version of the pdf, please do so.
%------------------------------------------------------------------------------

To accommodate highly-deformed configurations with high-energy HO excitations together with essential mixing of
low-energy excitations, typical SA-NCSM calculations  span the entire
(complete) space up to a given $N^{\bot}_{\max}$, while beyond this,
calculations include only selected many-nucleon basis states limited by the
$N_{\max}$ cutoff~\cite{DytrychLMCDVL_PRL13, DytrychHLDMVLO15, Dytrych12CPRC016}. 
We adopt a notation where an SA-NCSM
model space of ``$\langle N^{\bot}_{\max} \rangle N_{\max}$'' includes all the
configurations up through $N^{\bot}_{\max}$  and a restricted subspace beyond  $N^{\bot}_{\max}$ up through $N_{\max}$. 
%When we quote only the $N_{\max}$ value, it is understood that the space is complete through that $N_{\max}$ (for example $N_{\max}=8=N^{\bot}_{\max}$).
%We then specify the restrictions employed for each set of results where there is an SA-NCSM selection.
%
For example, the selections of basis states employed in this paper for a  $\langle 6 \rangle8$ model space are illustrated for $^{12}$C and $J=0$ states\footnote{Lists of the resulting configurations for other $J$-values are available upon request.} in Fig. \ref{Fig:12C_ space_selection}.
The $\langle 6 \rangle8$ space includes all configurations  up through 6\ho~  with the additional restriction of the 8\ho~ subspace to only selected deformations as well as selected  proton spin, neutron spin, and total spin values as shown in Fig. \ref{Fig:12C_ space_selection}. Namely, $(S_p, S_n, S) = (0, 0, 0),\, (0, 1, 1),\, (1, 0, 1),\, (1, 1, 0),\, (1, 1, 1)$, and $(1, 1, 2)$. A similar selection pattern is employed for $J=1,\,2$ and $J=4$.   The selection is based on high-deformation and low-spin dominance, along with symplectic \SpR{3} excitations thereof. Hence, configurations of largest deformation (typically, large $\lambda$ and $\mu$) and lowest spin values  are included first, that is, the lower right corner of the 8\ho~ subspace in Fig. \ref{Fig:12C_ space_selection}. The mathematical prescription for retaining states in these model spaces, together with the associated quantitative cutoff, 
is based on equations and conditions expressed within the \SU{3} and \SpR{3} groups  \cite{Dytrych12CPRC016} that are beyond the scope of this presentation.  
%We note, however, that the configurations represented in Fig.  \ref{Fig:12C_ space_selection} display the complete selections for $J=0$ states in $^{12}$C at $N_{\rm max} = 8$.
%is detailed in Ref.

A very important feature of the SA-NCSM is that any SA-NCSM selected model space of $\SU{3}\times\SU{2}_S$ irreps, that is, a space
spanned by all configurations carrying a fixed set of $S_{p}S_{n}S$ and
$N(\lambda\,\mu)$ quantum numbers, permits exact factorization of the
center-of-mass motion \cite{Verhaar60}. As a result, a reduced model space $\langle
N_{\max}^{\bot}\rangle N_{\max}$, defined by a set of physically relevant
$(\lambda\,\mu)$ and important intrinsic spins, yields eigenfunctions that exactly factorize into a product of
intrinsic and center-of-mass components. With the help of a Lagrange multiplier term, the wavefunctions, and associated observables, for states of interest (those free of center-of-mass excitation) are generated efficiently.
%are thus guaranteed to be free of spurious  center-of-mass motion.
%with the center-of-mass in the HO ground state.

%%%%%%%%%%%%%%%%%%%%%%%%%%%%%%%%%%%%%%%%%%%%%%%%%%%%%%%%%%%%%%%%%%%%%%%%%%%%%%%
%\subsection{$^{12}$C Properties Evaluated}
%%%%%%%%%%%%%%%%%%%%%%%%%%%%%%%%%%%%%%%%%%%%%%%%%%%%%%%%%%%%%%%%%%%%%%%%%%%%%%%

%%%%%%%%%%%%%%%%%%%%%%%%%%%%%%%%%%%%%%%%%%%%%%%%%%%%%%%%%%%%%%%%%%%%%%%%%%%%%%%
\section{$^{12}$C Properties and Model-space Selection Considerations\label{sec:C12properties}}
%%%%%%%%%%%%%%%%%%%%%%%%%%%%%%%%%%%%%%%%%%%%%%%%%%%%%%%%%%%%%%%%%%%%%%%%%%%%%%%
%Our goal here is to compare the SA-NCSM results with selected symmetry truncations to the exact NCFC results with the same $N_{\rm max}$ truncation.
%For this application, 
To examine the applicability of the SA-NCSM,
we select the ground state and eight low-lying positive-parity excited states
%7 states 
of $^{12}$C. Namely, three $0^+$ states,  three excited $2^+$ states, an excited $4^+$ state, as well as two excited $1^+$ states.
We evaluate and compare results for the binding energy, excitation energies, the ground-state point-particle matter rms radius,
%for the mass, neutrons and protons, 
the electric quadrupole moment of the $2^+_1$ and $4^+_1$ states  as well as the $B(E2; 2_1^+ \rightarrow 0^+_{\rm gs})$, the $B(M1; 1_1^+ \rightarrow 0^+_{\rm gs})$  transition strengths to the ground state and the magnetic dipole moment of the first excited $J=1$, $T=1$ state. 
%(UPDATE THIS IF WE DECIDE TO INCLUDE THE  $4^+_1$ STATE).
%\begin{widetext}
%------------------------------------------------------------------------------
% Table I: Binding energies
%------------------------------------------------------------------------------
\begin{table}[th]
\caption{
$^{12}$C observables  for selected $N_{\rm max}$ values, namely, binding energy
BE (MeV),  excitation energies $E$ (MeV), electric quadrupole moments $Q$ (e
fm$^2$) and magnetic dipole moment $\mu$ ($\mu_N$),  as well as $B(E2)$ ($e^2$
fm$^4$) and $B(M1)$ ($\mu_N^2$) reduced transition strengths, together with
point-particle matter rms radius  $r_m$ (fm). The observables are calculated
for \ho=20 MeV  using the bare JISP16 interaction and compared to the
experiment (``Expt.")~\cite{12CExpt}. The SA-NCSM  results are obtained in a reduced $\langle
6 \rangle8$ model space  with a complete space up to 6\ho~(``B" selection). 
Comparing MFDn and LSU3shell results up through $N_{\rm max}=6$ shows they are the same, to within the quoted precision.
%%Tomas, please confirm the use of Nmax=8  -- are Nmax=8 results available with LSU3shell, or did some of the results presented in the Table came from MFDn (if yes, please use Nmax=6)?
} 
%\label{Energies}
\begin{center}
    \begin{tabular}{l|lllll}
      \hline
      \hline
      & & \multicolumn{3}{c}{$N_{\rm max}$} &\\
        & Expt. & $4$ &  $6$ & $8$   & $\langle 6 \rangle$8-B\\
      \hline 
BE	&	92.162	&	72.654	&	82.192	&	87.902	&	85.951	\\
$E_{2^+_1}$	&	4.439	&	6.415	&	5.356	&	4.685	&	4.644	\\
$E_{1^+_1}$	&	12.71	&	17.024	&	15.307	&	14.161	&	14.199	\\
$E_{4^+_1}$	&	14.083	&	20.071	&	17.854	&	16.255	&	16.324	\\
       $Q_{2^+_1}$  	&   +6(3)	&      3.316      	&        3.546 	&       3.741    	&    	 3.735	 \\
       $Q_{4^+_1}$  	&   N/A	&      4.285      	&        4.597 	&       4.864    	&    	 4.845	 \\
       $\mu_{1^+_1}$  &   N/A      &      0.948        &        0.876      &      0.848       &       0.839       \\
       $B(E2; 2_1^+ \rightarrow 0^+_{\rm gs})$   
       				&  7.59(42)	& 2.723  		&    3.051    	&   3.342            &   3.301        \\
       $B(M1; 1_1^+ \rightarrow 0^+_{\rm gs})$   
       				&  0.0145(21) 	& 0.028    		&    0.018    	&   0.013            &    0.012    \\
       $r_m (0^+_{\rm gs})$  		& 2.43(2)\footnote{Ref. \cite{Tanihata85}}		& 1.996  		&      1.995   	&   2.003 		& 	2.005	\\
      \hline
      \hline
    \end{tabular}
\end{center}
\label{12C_observables}
\end{table}
%------------------------------------------------------------------------------
%\end{widetext}

We aim to study here model-space selection considerations in the SA-NCSM by
comparing results to the NCSM calculations for the same $N_{\rm max}$ cutoff.
We consider two SU(3)-based selection schemes, determined by symmetry
considerations, namely, ``$\langle 6 \rangle8$-A" (a smaller set of basis
states, Fig. \ref{Fig:12C_ space_selection}, filled circles, light yellow colors) and ``$\langle 6 \rangle8$-B'' (a
larger set of basis states, Fig. \ref{Fig:12C_ space_selection}, filled circles), with dimensions listed in Table
\ref{Table:NNZ_12C} and discussed in Sec. \ref{sec:complexity}.  
%Here, we briefly mention that the dimensions and, more importantly, the number of non-vanishing matrix elements (nnz) of these SA-NCSM model spaces are seen to be much smaller than the complete {\it J}-scheme dimensions and nnz also presented in Table \ref{Table:NNZ_12C}. 

\begin{figure}[th]
\begin{center}
\includegraphics[width=0.48\textwidth]{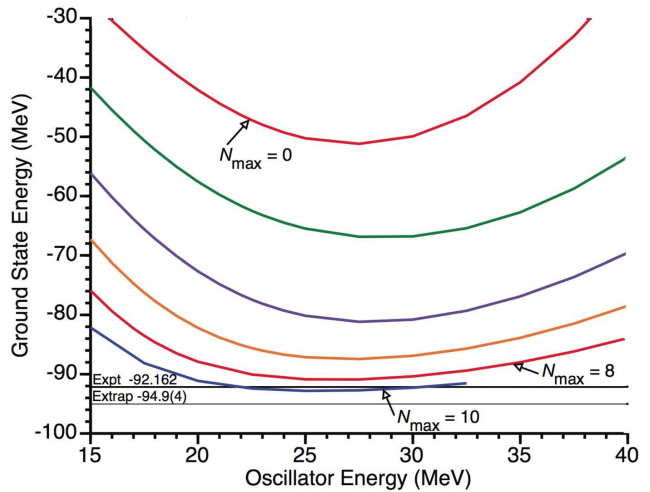}
\caption{Ground state energy of $^{12}$C as a function of $\hbar\Omega$ for a sequence
of $N_{\rm max}$ cutoffs 
%for complete spaces and  
with the JISP16 $NN$ interaction. 
%Similar patterns are obtained for the excited states of $^{12}$C.  
The quoted NCFC extrapolated energy is obtained using extrapolation method ``A" of reference \cite{Maris09_NCFCa}.  }
\label{Fig:12C_gs_vs_hw}
\end{center}
\end{figure}

%%%%%%%%%%%%%%%%%%%%%%%%%%%%%%%%%%%%%%%%%%%%%%%%%%%%%%%%%%%%%%%%%%%%%%%%%%%%%%%
%\subsection{Energies of Selected States}
%%%%%%%%%%%%%%%%%%%%%%%%%%%%%%%%%%%%%%%%%%%%%%%%%%%%%%%%%%%%%%%%%%%%%%%%%%%%%%%
We present the $gs$ energy of $^{12}$C obtained with {\tt MFDn} in the $M$-scheme basis using the bare JISP16 interaction 
%for complete spaces 
in Fig. \ref{Fig:12C_gs_vs_hw} through a sequence of $N_{\rm max}$ truncations and as a function of $\hbar\Omega$.
The energy converges uniformly from above as expected with increasing $N_{\rm max}$ 
and the curves become increasingly independent
of $\hbar\Omega$ (i.e. flatter) with increasing $N_{\rm max}$.  Both of these features are signals of $gs$ energy convergence.
We obtain significant increases in binding with each increment in $N_{\rm max}$.  The experimental result is indicated as a horizontal line.  For completeness, we also show
%We present 
the NCFC extrapolated result (extrapolation method ``A" of Ref. \cite{Maris09_NCFCa}) based on the results through $N_{\rm max}=10$ \cite{Maris09_NCFCc, Maris09_NCFCd}. 
%
%------------------------------------------------------------------------------
% Fig: Results as a function of basis space parameters
%------------------------------------------------------------------------------
%------------------------------------------------------------------------------

In Table \ref{12C_observables}, we compare the results for $^{12}$C observables
at $\hbar\Omega=20$ MeV obtained with  {\tt MFDn} (see also \cite{MarisHITES12}) and {\tt LSU3shell} at the
SU(3) selection scheme ``$\langle 6 \rangle$8-B" mentioned above and discussed
in Sec. IV below.  
Table \ref{12C_observables} reveals the remarkable result
that the binding energy in $\langle 6 \rangle$8-B, with only 0.67\% of the corresponding complete model space,  reproduces 98\% of the $N_{\rm max}=8$ NCSM binding
energy. In fact, the $\langle 4 \rangle8$-B model space already reproduces 96\%
of this observable.  This points to the fact that complete 0\ho, 2\ho, and
4\ho~ model spaces ($N^{\bot}_{\rm max}=4$) and only selected physically
relevant 6\ho~ and 8\ho~ basis states suffice to capture most of the physics
that governs the $^{12}$C ground state. Further improvements are observed in the
case of $\langle 6 \rangle8$-B, but these improvements are less significant for
various observables. In Table \ref{12C_observables}, we also provide the experimental counterparts and note that, in general,
meaningful comparisons between computations and experiment require both 
%improved
convergence with $N_{\rm max}$ and $N^{\bot}_{\rm max}$, along with quantification of numerical uncertainties due to basis truncations.
%the main source of differences between the current results and experiment is probably the lack of complete convergence,  which would require even larger model spaces. To achieve accurate descriptions, convergence of SA-NCSM results needs to be assured with increasing  $N_{\rm max}$ and $N^{\bot}_{\rm max}$.

%------------------------------------------------------------------------------
% Fig: Binding energies of the selected low-lying states of 12C: vs Nmax
%------------------------------------------------------------------------------
%
\begin{figure}[th]
\includegraphics*[width=1.0\columnwidth]{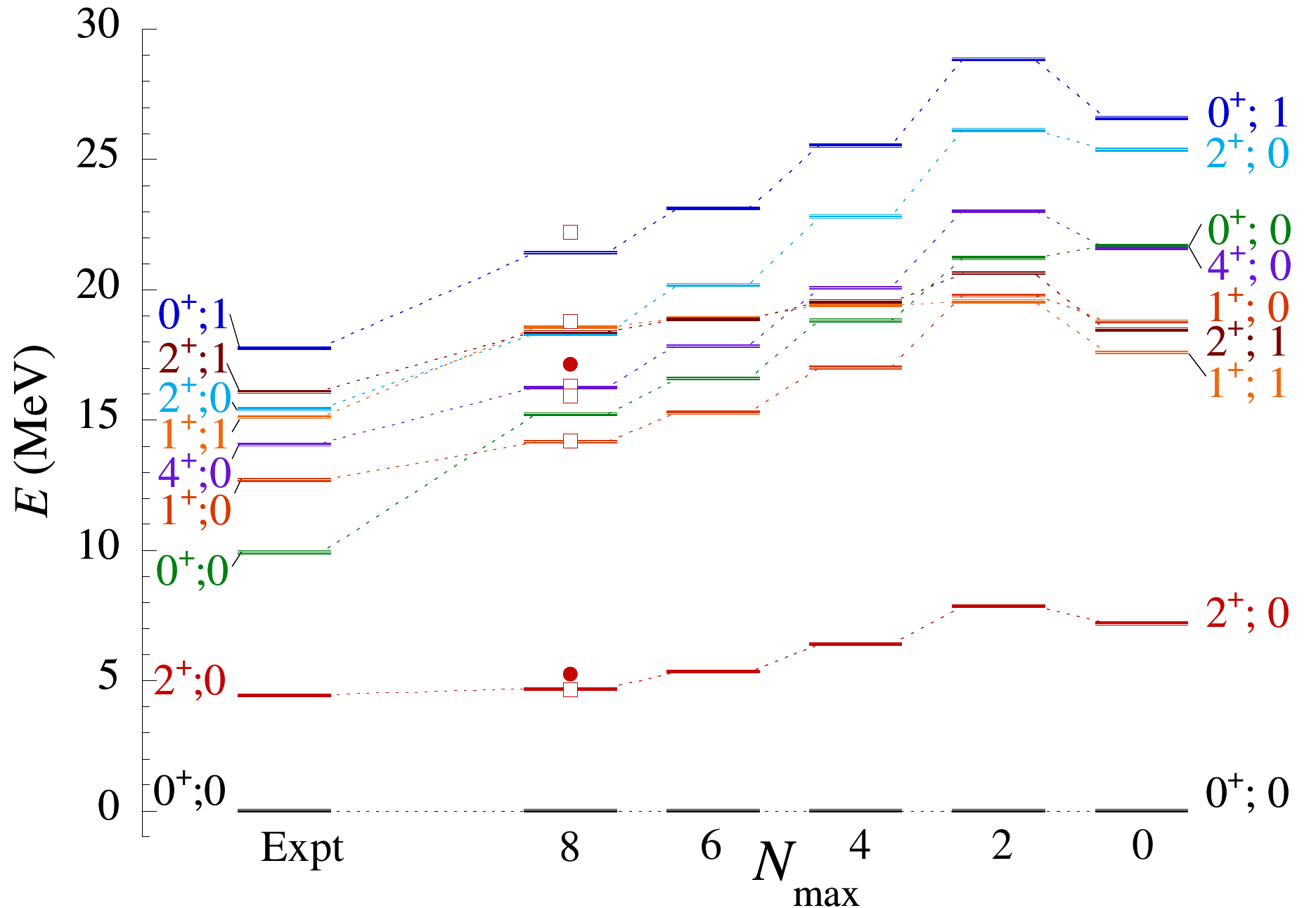} 
\caption{ 
$^{12}$C eigenenergies obtained with the  {\tt MFDn} and the bare JISP16 interaction for $\hbar\Omega=20$ MeV as a function of $N_{\rm max}$.
The corresponding SA-NCSM results for $\langle 6 \rangle8$ are shown for the ``A" (filled circles) and ``B" (empty squares) selection schemes; the order of the empty squares is as follows, from bottom to top, $2^+_1$ ($2^+;0$), $1^+_1$ ($1^+;0$), $0^+_3$ ($0^+;0$), $4^+_1$ ($4^+;0$), $2^+_2$ ($2^+;0$), and $0^+_4$ ($0^+;1$). Note that the 7.65-MeV $0^+_2$ state (the so-called Hoyle state) is not included in the plot.
%Model space  dimensions  for the SA-NCSM calculations, 
%%given as a percentage of the $N_{\rm max}=8$ complete model space,  are as following: for ``A", 0.47\% ($J=0$), 1.72\% ($J=2$), %and 1.58\% ($J=4$); and for ``B", 0.67\% ($J=0$), 2.75\% %%%($J=2$), 3.42\% ($J=4$), and 2.43\% ($J=1$).
%\blue{as well as the number of non-vanishing matrix many-body matrix elements, are listed in Table \ref{Table:NNZ_12C}
%and compared with the corresponding complete $N_{\rm max}=8$ {\it J}-scheme values.}
}
\label{12C_excEn_vsNmax}
\end{figure}
\begin{figure}[th]
\includegraphics*[width=0.9\columnwidth]{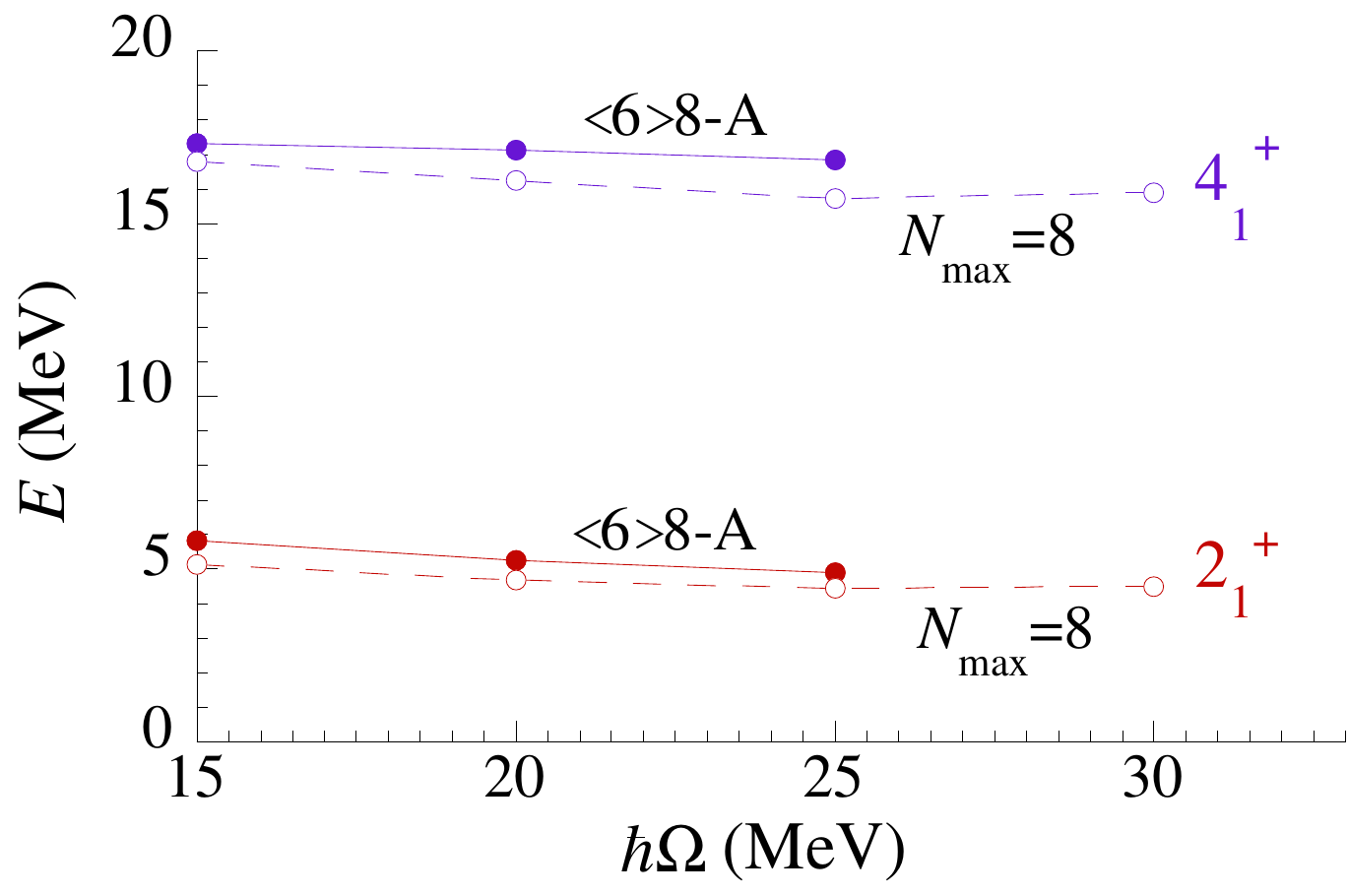}
\caption{ 
$^{12}$C eigenenergies as a function of \ho~ obtained with the SA-NCSM and the bare JISP16 interaction  for the $\langle 6 \rangle8$-A selection scheme (filled circles, solid lines), and compared to the complete $N_{\rm max}=8$ model space (open circles, dashed lines). 
}
\label{12C_excEn_vshw}
\end{figure}

To study convergence of excitation energies, we  present the low-lying spectrum
of  $^{12}$C at $\hbar\Omega=20$ MeV as a function of $N_{\rm max}$ (Fig.
\ref{12C_excEn_vsNmax}). The outcome reveals that with increasing $N_{\rm
max}$, the difference between the theoretical and experimental excitation energies
is seen to decrease.
%(the largest deviation at $N_{\max}=8$ is observed for the 10.3-MeV $0^{+}$ state).  
The $\langle 6 \rangle$8-B SA-NCSM calculations  for the
isospin-zero $2^+_1$, $1^+_1$, and $4^+_1$ excited states (Table
\ref{12C_observables} and Fig. \ref{12C_excEn_vsNmax}, empty squares) deviate
from the corresponding $N_{\rm max}=8$ results only by 0.9\%, 0.3\%, and 0.4\%,
respectively.  Higher-lying states, such as $0^+_3$, $2^+_2$, and $0^+_4$, exhibit a slightly larger deviation, with the largest difference observed for $0^+_4$, which is still only about 800 keV. These  states are found to lie remarkably close to the
complete-space counterparts even when the smaller $\langle 6 \rangle$8-A
SA-NCSM model space is utilized (see Fig. \ref{12C_excEn_vsNmax} for $2^+_1$ and $4^+_1$, filled circles).
In addition, we compare excitation energies  as a function of $\hbar\Omega$ in
the range of $15$ to $25$ MeV. In this region, we find that restricted SA-NCSM
model spaces yield excitation energies that change only slightly with \ho, e.g.,  for
the $2^+_1$ and $4^+_1$ $T=0$ states, the energies change by less than 1 MeV through a
change of 10 MeV in the \ho~oscillator energy (Fig. \ref{12C_excEn_vshw}).  
%------------------------------------------------------------------------------
% Fig: Binding energies of the selected low-lying states of 12C: vs hw
%------------------------------------------------------------------------------
%
%

%%%%%%%%%%%%%%%%%%%%%%%%%%%%%%%%%%%%%%%%%%%%%%%%%%%%%%%%%%%%%%%%%%%%%%%%%%%%%%%
%\subsection{Electric Quadrupole Moments, $B(E2)$, $B(M1)$, and Radii }
%%%%%%%%%%%%%%%%%%%%%%%%%%%%%%%%%%%%%%%%%%%%%%%%%%%%%%%%%%%%%%%%%%%%%%%%%%%%%%%
Table \ref{12C_observables} also presents restricted-space and complete-space
results for the electric quadrupole moment of the first excited $2^+$ and $4^+$
states of $^{12}$C,  the reduced probability for the $E2$ transition  of
$2^+_1$ and $M1$ transition of $1^+_1$  to the ground state, as well as the
point-nucleon matter rms radius of the ground state. These observables are
remarkably well reproduced by the  SA-NCSM calculations in the $\langle 6
\rangle8$-B space.  Similar results are obtained already in the $\langle 4
\rangle$8-B model space.  For example, $B(E2;2_1^+ \rightarrow 0^+_{\rm
gs})=3.339$ $e^2$fm$^4$, which represents $99.9\%$ of the complete space
result.  The smaller $\langle 6 \rangle8$-A model space yields,
$Q_{2^+_1}=3.712$ $e$fm$^2$, $Q_{4^+_1}=4.826$ $e$fm$^2$,  $B(E2; 2_1^+
\rightarrow 0^+_{\rm gs})=3.289$ $e^2$fm$^4$, and $r_m(0^+_{\rm gs})=2.007$ fm,
which already reproduce more than 99\% of the $N_{\rm max}=8$ NCSM results.
This indicates that these observables are not sensitive to the fine-tuning of
the selected space and only a few  configurations of the 8\ho~ subspace (shown
in Fig. \ref{Fig:12C_ space_selection}) appear sufficient for their accurate
description. 

In short, we show that SA-NCSM yields results in reduced \SU{3}-based model
spaces that practically coincide  with the ones obtained in the corresponding
complete space. The computational complexity associated with such calculations
is examined next along with exploration of memory savings achievable in larger
spaces using the examples of $^6$Li and $^{12}$C.

%%%%%%%%%%%%%%%%%%%%%%%%%%%%%%%%%%%%%%%%%%%%%%%%%%%%%%%%%%%%%%%%%%%%%%%%%%%%%%%
\section{Computational Complexity \label{sec:complexity}}
%%%%%%%%%%%%%%%%%%%%%%%%%%%%%%%%%%%%%%%%%%%%%%%%%%%%%%%%%%%%%%%%%%%%%%%%%%%%%%%
Issues governing the required computational resources include the model space dimension, the
number of non-vanishing many-body matrix elements and the computational effort required by those matrix elements \cite{MFDn1, MFDn2, MFDn3, MFDn4}. For the highly scalable algorithms we have developed and implemented, the computational resources can be viewed primarily in terms of memory and time requirements.
We now turn our attention to these computational complexity issues.

The nuclear many-body calculations we address here involve evaluating the
Hamiltonian in a selected basis representation and solving the resulting matrix
for a small set (typically less than 20) of its low-lying eigenvalues and
eigenvectors. To solve the large sparse matrix for its eigenvalues and
eigenvectors on a massively parallel architecture is recognized as
computationally hard.  The challenge in nuclear physics is compounded by the
strong inter-nucleon interactions referred to in Eq.~(\ref{intH}) \cite{MFDn1}
that induce significant short- and intermediate-range inter-nucleon
correlations.  The model space must therefore be sufficiently large to account
for these correlations and, at the same time, account for the long-range
``tails'' of the nuclear wavefunctions.  The need for accurate representations
of the short, intermediate and long-range features of the wavefunctions drives
the computational resources in different ways for the methods we employ.   
For example, the need for particular excited states (such as members of a rotational band) that lie in a region of high level density motivates the fixed-$J$ basis.

Each method we employ involves  a transformation of stored $NN$ (and possibly
$NNN$) matrix elements to the chosen many-nucleon basis representation.  This
process begins with initial interaction matrix elements stored in the HO basis
that are reduced matrix elements, i.e. matrix elements in a basis
coupled to total angular momentum $J$ (and sometimes total isospin $T$), and accommodate
%Separate initial data sets are employed to manage 
charge-dependent interactions. 
Currently, this initial interaction scheme is employed for both $NN$ and $NNN$ interactions but
only $NN$ interactions are employed here. 

The transformation of these initial interaction data sets to the representation
of the basis of choice involves ``recoupling transformations'' that may be
viewed as multiplications of 3 non-square matrices (see Ref. \cite{Oryspayev}
for a graphical illustration). Our different choices of many-nucleon basis (e.g.,
$M$-scheme basis, {\it J}-scheme basis, and SU(3)-scheme basis) involve
different recoupling transformations (see Ref. \cite{LauneyDDSD15} for the SU(3) scheme).  While for the NCSM, this transformation
is part of the {\tt MFDn} code, the SA-NCSM invokes a separate algorithm, which
transforms each interaction set only once  and stores it for use by the {\tt
LSU3shell}.  Improving the efficiency of the recoupling transformations
themselves and reducing their memory footprints are subjects of intensive
ongoing research and involves, among other issues, efficient exploitation of
computer architectures such as Graphics Processing Units. We will
limit our considerations here to the computationally demanding evaluation of
the many-body Hamiltonian and computation of the corresponding eigenvalues and
eigenvectors.

\begin{figure}[th]
\begin{center}
\includegraphics[width=\columnwidth]{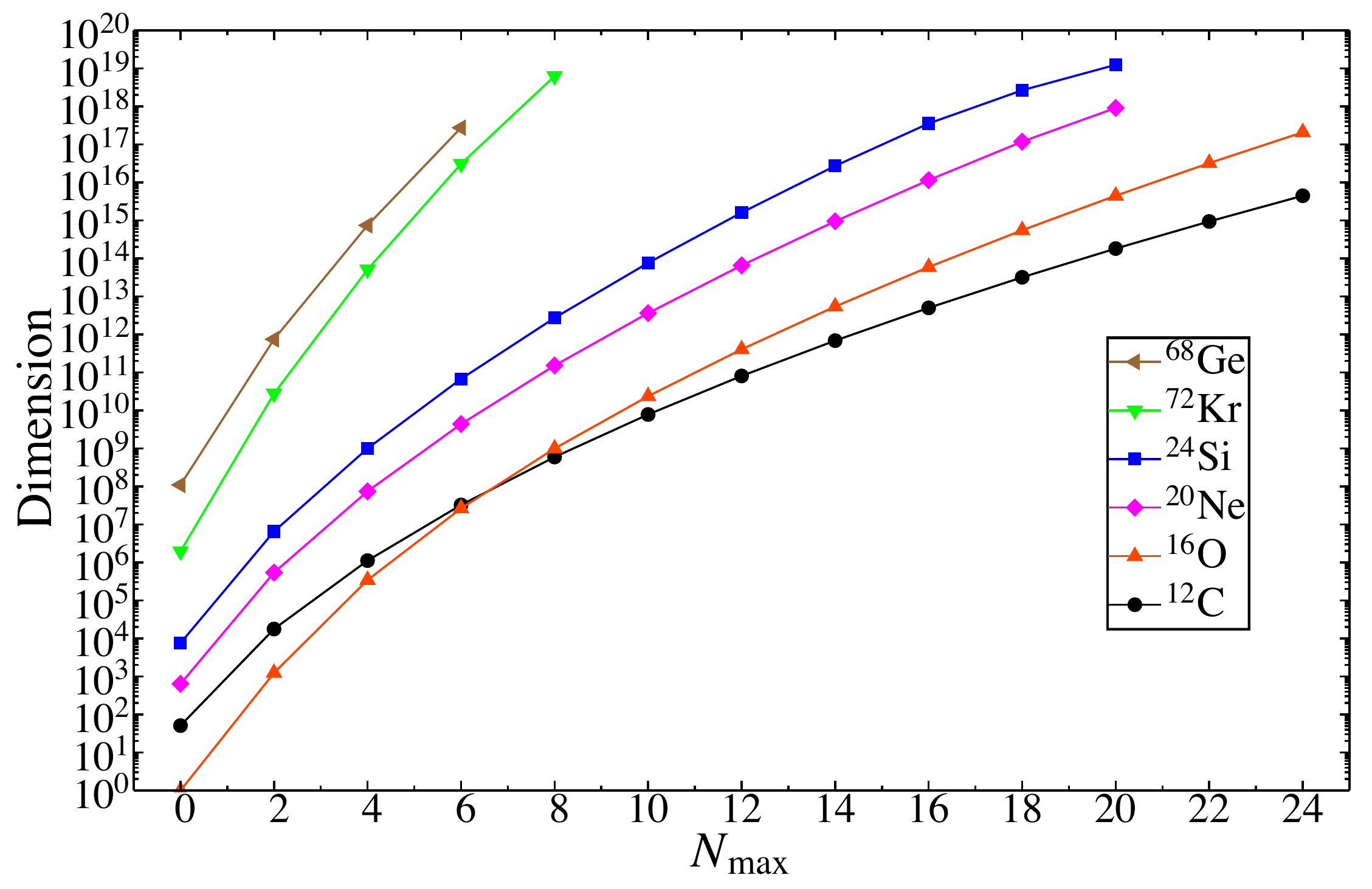}
\end{center}
\caption{
\label{Fig:Nmax-Mscheme}
Dimensions of the $M$-scheme (with $M=0$) natural parity model spaces for
the even-even $N=Z$ nuclei with $N_{\rm max}$ truncation.  The sequence
of dimensions for unnatural parity states (odd values of $N_{\rm max}$) lie
intermediate to the neighboring natural parity dimensions.
}
\end{figure}

Figure \ref{Fig:Nmax-Mscheme} presents model space dimensions (size of the
many-body Hamiltonian matrix) in the $M$ scheme over a range of $N_{\rm
max}$ values for even-even $N=Z$ nuclei.  The figure illustrates the
dramatic increase of matrix dimensions with increasing $N_{\rm max}$ and
increasing atomic number $A$. For example, the corresponding $^{12}$C data points
show the size of the $M$-scheme basis used at each  $N_{\rm max}$ to
produce the results of Fig. \ref{Fig:12C_gs_vs_hw}.  For modern realistic $NN$
and $NNN$ interactions, we have found for $p$-shell nuclei that $N_{\rm max}$
values of 8 and above are desirable for achieving results approaching
convergence~\cite{Maris09_NCFCa, Maris09_NCFCc,Maris09_NCFCd,Coon:2012ab,Furnstahl:2012qg,Maris2013aa_JPV,Jurgenson2013aa_JPV}.
This increase in dimensionality and associated increase in the number of non-zero many-body matrix elements motivate the
search for model spaces with the aim of reducing the computational complexity
as also pursued by other efforts mentioned earlier~\cite{RothN07,Otuska_MCSM1,Otuska_MCSM2,Otuska_MCSM3}.

\begin{figure}[th]
\begin{center}
\includegraphics[scale=0.38]{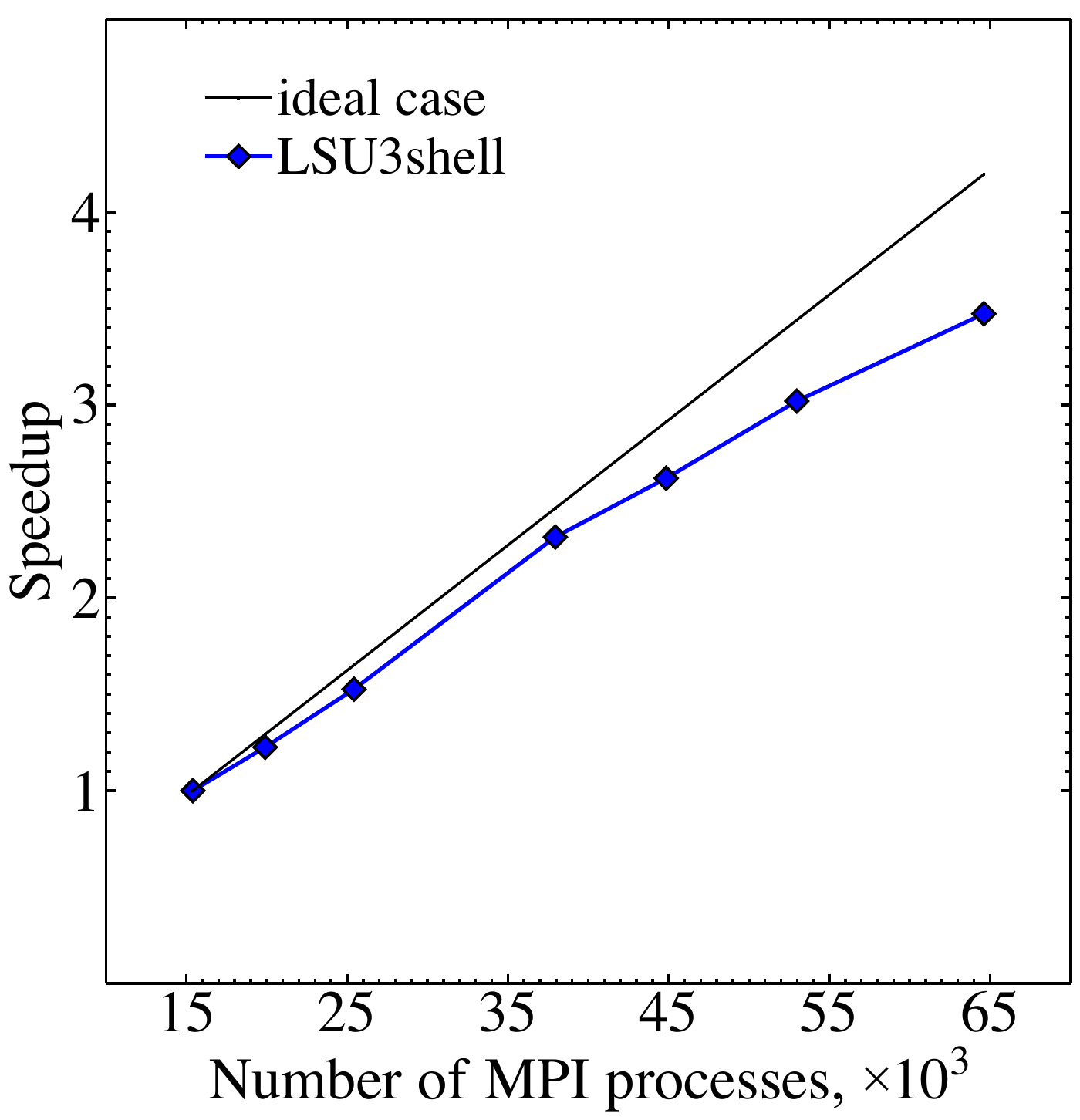}\\
(a)\\
\includegraphics[scale=0.38]{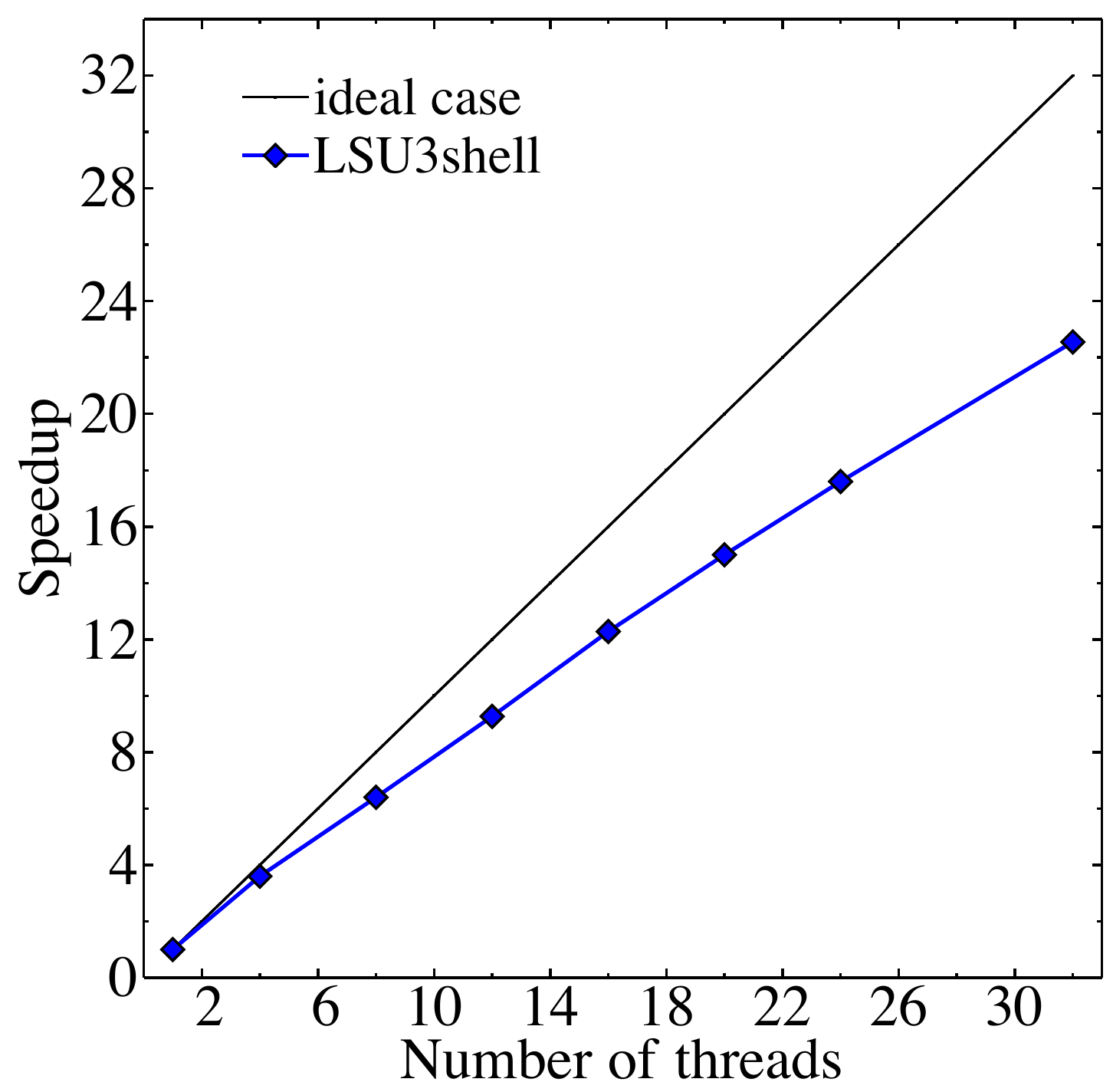} \\
(b)\\
\caption[]{\label{fig:StrongScaling}
(a) Strong scaling of the Hamiltonian matrix construction for $J=0$
	states of $^{24}$Mg (blue
	curve, $\dim=0.98\times 10^{6}$) as a function of the number of MPI processes
	(in units of thousands).
	(b) Strong scaling of \texttt{LSU3shell} on a single Cray XE6 compute node
	as a function of the number of OpenMP threads. (Scaling results are obtained on the Blue Waters system.)
}
\end{center}
\end{figure}

\noindent
{\bf SA-NCSM runtime and number of single-shell configurations. -- } 
Computing the nuclear many-body Hamiltonian matrix in an \SU{3}
symmetry-adapted basis within the SA-NCSM framework represents the most
computationally intensive task -- one that typically takes over 95\% of the
total CPU time.  Fortunately, this task can be considered a perfectly parallel
problem and endows \texttt{LSU3shell} code with good scalability to hundreds of
thousands of cores and possibly even beyond (Fig.~\ref{fig:StrongScaling}).
Each MPI process is assigned a submatrix of the Hamiltonian and invokes OpenMP
threads to share the workload.  The communication network is used merely to
distribute data between collaborating processes during the initial set-up
phase.  No other communication between processes is needed, as each matrix
element of the many-body Hamiltonian matrix can be evaluated independently. \SU{3} basis
states are mapped to the MPI processes in a block round-robin fashion, where
the blocks are defined by similar \SU{3} structures using the same set of
\SU{3} coupling coefficients and reduced matrix elements needed to evaluate
Hamiltonian matrix elements, as described below.  This approach leads to a
uniform distribution of matrix elements and thus allows one to achieve a
reasonably good load balancing.

\begin{figure}[th]
\begin{center}
\includegraphics[width=\columnwidth]{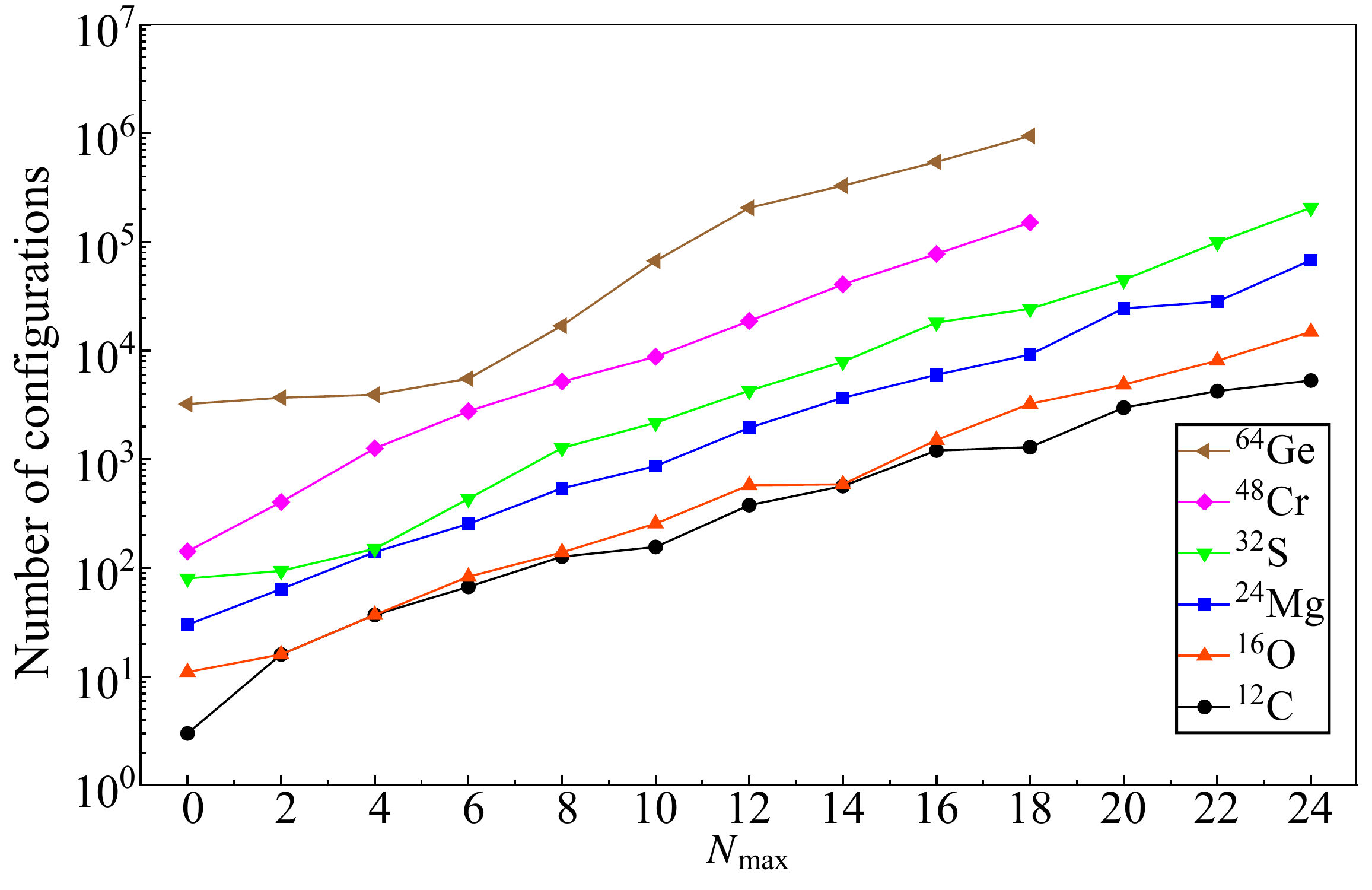}
\caption{Number of single-shell $\Un{\Omega_\eta}\times\SU{2}_{S}$ irreps, with $\Un{\Omega_\eta}\supset \SU{3}$, that
generate the SA-NCSM model space for even-even $N=Z$ nuclei
as a function of $N_{\rm max}$ cutoff.
 }
\label{Fig:dimensions_hws}
\end{center}
\end{figure}

The SA-NCSM implements fast methods for calculating matrix elements of
arbitrary operators in the symmetry-adapted basis.  This facilitates both the
evaluation of the Hamiltonian matrix elements and the use of the resulting
eigenvectors to evaluate other experimental observables.  The underlying
principle behind the SA-NCSM computational kernel is an \SU{3}-type
Wigner-Eckhart theorem, which factorizes interaction matrix elements into the
product of \SU{3} reduced matrix elements (RME)  and the associated \SU{3}
coupling coefficients.  To compute the Hamiltonian matrix elements,
\texttt{LSU3shell} adopts state-of-the-art group-theoretical
methods~\cite{DraayerLPL89}, implemented in C++ programming language, and
optimized Fortran numerical subroutines~\cite{AkiyamaD73} for computing
required \SU{3} coupling/recoupling coefficients.  As described above, the
SA-NCSM configurations are constructed as the inter-shell coupling of a set of
single-shell irreps of $\Un{\Omega_\eta}\times\SU{2}_{S}$ with $\Un{\Omega_\eta}\supset
\SU{3}$.  Therefore, all the multi-shell RME are constructed from a set of
single-shell RME computed in a configuration space of the single-shell irreps.
This reduces the number of key pieces of information required to the
single-shell RME, and these track with the number of the single-shell
$\Un{\Omega_\eta}\times\SU{2}_{S}$ irreps, with $\Un{\Omega_\eta}\supset \SU{3}$, that
represent building blocks of the SA-NCSM approach.  It is therefore significant
that their number grows slowly with the increasing nucleon number and
$N_{\max}$ cutoff as illustrated in Fig. \ref{Fig:dimensions_hws}, as this
allows these key pieces of information to be stored in CPU memory in the {\tt
LSU3shell} code.  
% For example, the largest number of configurations appearing in Fig. 7 is about one million (for $N_{\rm max} = 18$ and $^{64}$Ge) for which  the associated single-shell RMEs would fit comfortably within one GB.
%%Tomas, please confirm the last sentence above
%------------------------------------------------------------------------------
%------------------------------------------------------------------------------

An important feature of the SA-NCSM is that the symmetry-guided organization of
the model space allows the full space to be systematically down-selected to the
physically relevant subspace.  As a second illustrative  example, we consider
SA-NCSM results for $^6$Li low-lying states that, as shown in Ref.
\cite{DytrychLMCDVL_PRL13}, have achieved significantly reduced dimensions for
equivalent large shell-model spaces  without compromising the accuracy of the
{\it ab initio} NCSM approach (see Fig. 2 and Fig 3 of Ref.
\cite{DytrychLMCDVL_PRL13} for  $gs$ and excitation energies, $B(E2)$ values
and $Q$ electric quadrupole moments).  The runtime of the SA-NCSM code exhibits
a quadratic dependence on the number of  $(\lambda\,\mu)$ and  $(S_p S_n S)$
irreps -- there are $1.74\times10^{6}$ such irreps for the complete $N_{\max}$
= 12 model space of $^{6}$Li, while only 8.2\%, 8.3\%, 8.9\%, 12.7\%, and
30.6\% of these are retained when keeping all allowed irreps at
$N^{\bot}_{\max} = 2, 4, 6, 8,$ and $10$, respectively.  The net result is that
calculations in the $\langle 10\rangle 12$, $\langle 8\rangle 12$, \dots,
$\langle 2\rangle 12$ spaces require one to two orders of magnitude less
computational time than SA-NCSM calculations for the complete $N_{\max}=12$
space.

\begin{figure}[th]
\begin{center}
\includegraphics[width=1\columnwidth]{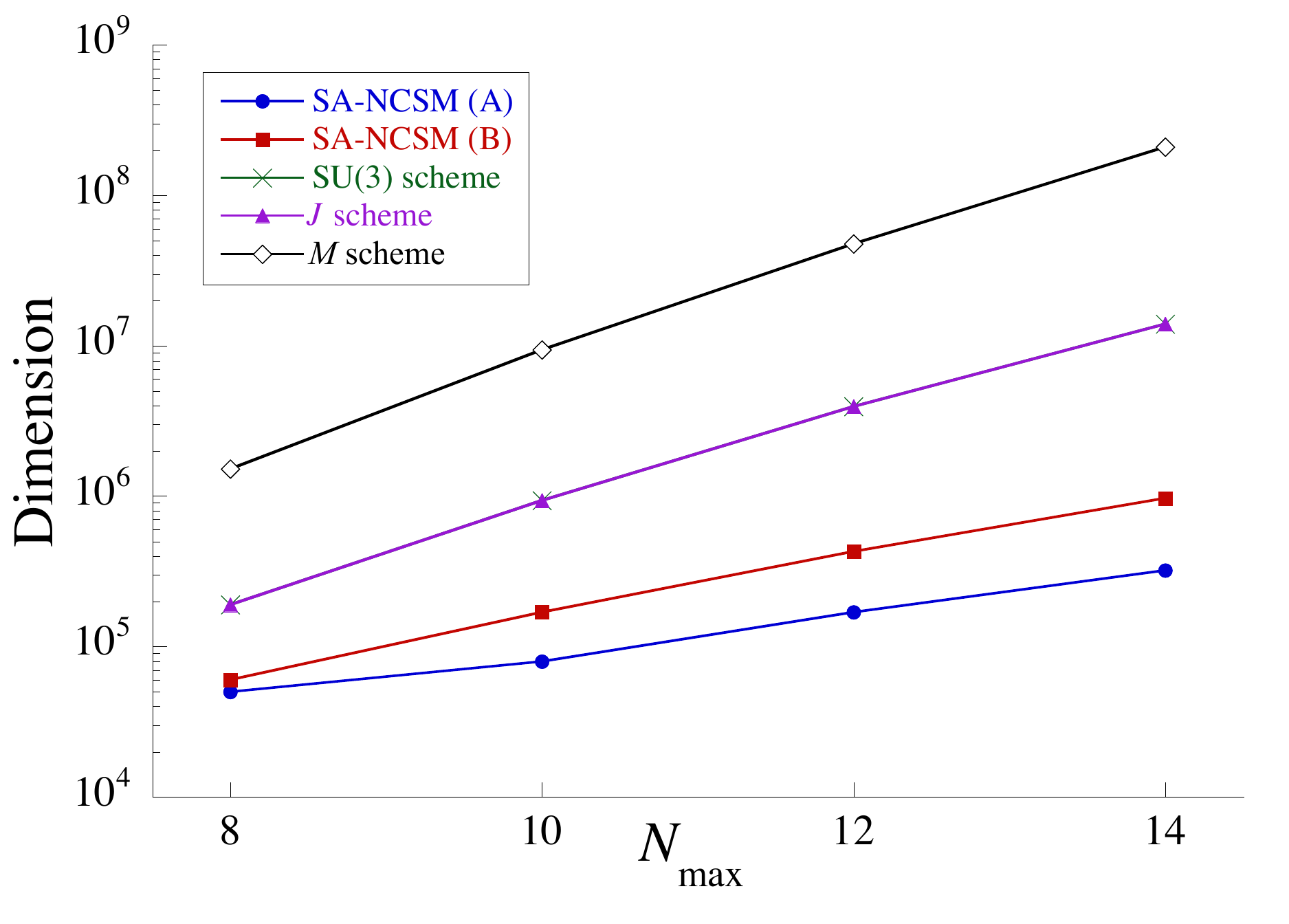}\\
(a)\\
\includegraphics[width= 1\columnwidth]{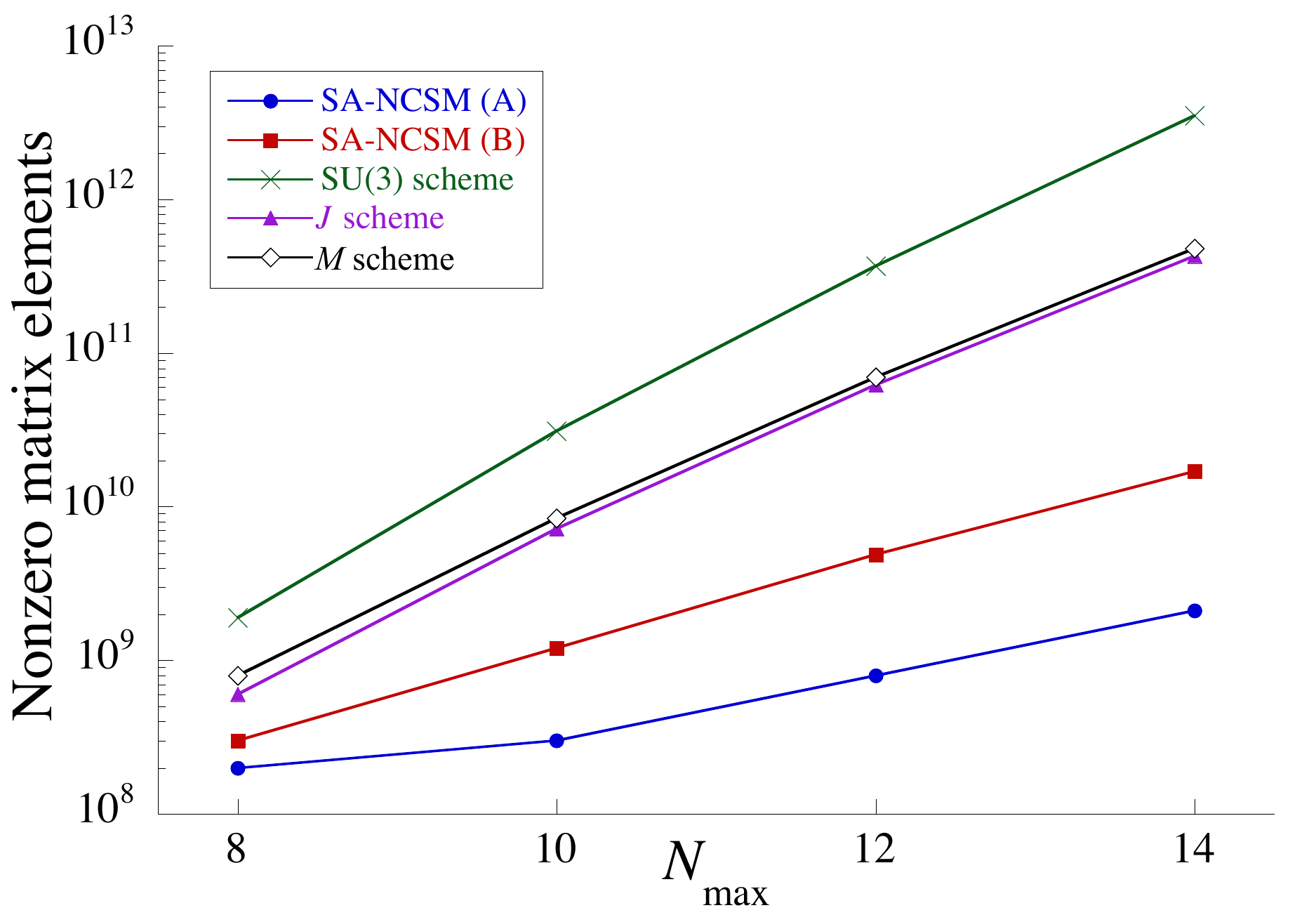}\\
(b)
\caption{
(a) Matrix dimensions  and (b) number of non-zero (nnz)
many-body matrix elements for $^{6}$Li and $J=1$  as functions of   
the $N_{\max}$ cutoff for selected \SU{3} model spaces and for the complete $N_{\rm max}$ space for the \SU{3}, $J$, and $M$ schemes. The dimension for the $J$ scheme coincides with that for the \SU{3} scheme;
the total $M$-scheme dimension is shown for all $J\ge 1$  (for $M=1$).  The nnz entries are quoted for an input $NN$ interaction.
 }
\label{Table:NNZ_6Li}
\end{center}
\end{figure}

\noindent
{\bf Number of non-vanishing many-body Hamiltonian matrix elements -- } A very
small fraction of the SA-NCSM runtime is devoted to the eigenvalue procedure.
Since {\tt MFDn} as well as the current implementation of {\tt LSU3shell},
which utilizes the Lanczos parallel eigensolver of {\tt MFDn},  store the
non-vanishing many-body Hamiltonian matrix elements, the number of these
elements in a specified basis dominates the storage requirements in our
applications.  For example, the number of the non-vanishing matrix elements in
the $\langle 6 \rangle8$-B space are comparable or up to an order of magnitude
greater than the $M$-scheme counterparts, as shown in Table \ref{Table:NNZ_12C}
 for $^{12}$C and in Fig. \ref{Table:NNZ_6Li} (see the $N_{\rm max}=8$ entries) for  $^{6}$Li  (observables obtained in the model spaces discussed here for  $^{6}$Li  are presented in Ref. \cite{DytrychLMCDVL_PRL13}).  The $^{12}$C
scenario probably realizes the complexity upper limit of the SU(3) scheme,
since it is actually an exception rather than a typical performance. The reason
is that the dominant deformation in $^{12}$C is oblate and this results in
relevant subspaces that usually consist of less deformed configurations of
comparatively  larger  dimensionality. A more typical performance is expected for a
nucleus like $^{6}$Li (Fig. \ref{Table:NNZ_6Li}) with low-lying states of a
prolate dominant deformation, which is the deformation favored by most nuclei.
%------------------------------------------------------------------------------
\begin{table}[th]
\caption{
Matrix dimensions (``dim", first row for each $J$) and number of non-vanishing 
many-body matrix elements (``nnz", second row for each $J)$
for $^{12}$C at selected values of total angular momentum and their dependence 
on the model space for $N_{\rm max}=8$.  
For reference, the total $M$-scheme dimension at $N_{\max}=8$ is
5.9$\times 10^8$  and the total number of non-vanishing matrix elements 
in this basis is 4.95$\times 10^{11}$. 
The dimension in the complete space is independent of the coupling scheme. 
%The column labelled ``$J$-scheme'' features results from a simple recoupling of the $M$-scheme basis.
The nnz entries are quoted for an input $NN$ interaction and, for the $J=1$ and $J=4$ $J$-scheme  entries, are estimates that should be accurate to better than 10\%.
}
\label{Table:NNZ_12C}
 \begin{center}
   \begin{tabular}{cllllll}
\hline
\hline
& & \multicolumn{2}{c}{$\langle 6 \rangle 8$} & & \multicolumn{2}{c}{complete space} \\ 
 \cline{3-4}  \cline{6-7}
$J$ & & \multicolumn{1}{c}{A}                      &    \multicolumn{1}{c}{B}                     &             & \multicolumn{1}{c}{SU(3)} & \multicolumn{1}{c}{$J$-scheme}  \\ \hline
\multirow{2}{*}{$J=0$} 
& dim &   2.8$\times 10^6$    &  4.0$\times 10^6$    &         & 1.9$\times 10^7$    &   1.9$\times 10^7$            \\
& nnz &   1.0$\times 10^{11}$ &  1.7$\times 10^{11}$ &         & 2.5$\times 10^{12}$ &   9.5$\times 10^{11}$   \\ 
\\
\multirow{2}{*}{$J=1$} 
& dim & \multicolumn{1}{c}{--}     &   1.4$\times 10^7$        &         &  5.4$\times 10^7$    & 5.4$\times 10^7$   \\
& nnz & \multicolumn{1}{c}{--}     &   1.9$\times 10^{12}$     &         &  2.0$\times 10^{13}$ & \multicolumn{1}{c}{$1.3 \times 10^{13}$}   \\ 
\\
\multirow{2}{*}{$J=2$} 
& dim & 1.0$\times 10^7$        &   1.6$\times 10^7$      &      &   7.9$\times 10^7$  & 7.9$\times 10^7$                    \\
& nnz & 1.2$\times 10^{12}$     &   1.6$\times 10^{12}$   &      &  2.2$\times 10^{13}$  & 1.7$\times 10^{13}$       \\ 
\\
\multirow{2}{*}{$J=4$} 
& dim & 9.4$\times 10^6$        &   2.0$\times 10^7$      &      &   9.1$\times 10^7$  & 9.1$\times 10^7$                    \\
& nnz & 1.7$\times 10^{12}$     &   3.9$\times 10^{12}$   &      &  3.9$\times 10^{13}$  &    \multicolumn{1}{c}{$3.1 \times 10^{13}$}   \\ 
\hline
\hline
\end{tabular}
 \end{center}
\end{table}

Above all, as shown in  Fig.  \ref{Table:NNZ_6Li}, an important feature of the
SA-NCSM is that the number of the non-vanishing matrix elements shows a slower
increase with $N_{\rm max}$, as compared to the {\it J} and $M$ schemes.
Clearly, this number for the `B' selection is $50\%$, $17\%$, $8\%$, and $4\%$ of the corresponding {\it
J}-scheme calculations in the complete $N_{\rm max}$ = 8, 10, 12, and 14 basis (it
is 38\%, 14\%, $7\%$, and $3.5\%$ compared to the $M$-scheme basis). This indicates
that  model spaces that are currently accessible on modern-day computer
architectures can reach higher $N_{\rm max}$ within the SA-NCSM framework
(e.g., $\langle 2 \rangle6$ for $^{24}$Si \cite{DraayerDLDL15}).  Such an
efficient computational scaling with $N_{\rm max}$ is essential for the
applicability of the SA-NCSM to larger model spaces needed to address
largely deformed and spatial cluster configurations.
%------------------------------------------------------------------------------
%------------------------------------------------------------------------------

\noindent
{\bf Dimension  of model space (Hamiltonian matrix) -- } 
Analysis of results reveals that the SA-NCSM accurately reproduces
corresponding complete-space results by using only a fraction of the model
space. For example, the  $\langle 6 \rangle$8-A ($\langle 6 \rangle$8-B)
selection for $^{12}$C  (Table \ref{Table:NNZ_12C}) produces dimensions that
range from $10\%$ to $15\%$ (from $21\%$ to $27\%$) of the corresponding {\it
J}-scheme NCSM basis dimensions as well as  from $0.5\%$ to $1.6\%$ (from
$0.7\%$ to $3.4\%$) of the corresponding $N_{\rm max}=8$ $M$-scheme NCSM
basis dimensions (similar fractions are found for $^{6}$Li). The smaller number of basis states needed in the SA-NCSM for a successful description of these nuclear states points to a faster rate of convergence achieved in the symmetry-adapted framework, as shown in Refs. \cite{DytrychLMCDVL_PRL13} and \cite{Dytrych12CPRC016}.  The drastically
reduced model spaces that are found sufficient to describe nuclear properties
of $p$-shell nuclei confirm the physical relevance of the \SU{3} basis for
nuclear modeling. 

\vspace{0.1in}
In short, we have shown that solutions of the SA-NCSM with symmetry-dictated
model-space selections  retain accuracy and are highly scalable on massively
parallel architectures. They also  require computations for significantly
reduced model-space dimensions and nnz compared with the fixed-$J$
basis of NCSM while enhancing access to converged nuclear properties.

%%%%%%%%%%%%%%%%%%%%%%%%%%%%%%%%%%%%%%%%%%%%%%%%%%%%%%%%%%%%%%%%%%%%%%%%%%%%%%%
\section{Summary}
%%%%%%%%%%%%%%%%%%%%%%%%%%%%%%%%%%%%%%%%%%%%%%%%%%%%%%%%%%%%%%%%%%%%%%%%%%%%%%%

We have demonstrated the efficacy of the \SU{3}-scheme that utilizes symmetries
to reduce the dimensionality of the model space and, more importantly, the
number of non-zero many-body matrix elements through a very structured
selection of the basis states to physically relevant subspaces without
compromising the accuracy of the {\it ab initio} NCSM approach. 

While applications of the NCSM and SA-NCSM in complete model spaces necessarily
yield identical results (indeed, this is how the  SA-NCSM
implementation was validated), the intended use of the two approaches is
complementary.  The NCSM offers the possibility to produce complete-space
results for light nuclei; the SA-NCSM is intended for applications beyond the
lightest nuclei and to cluster formation where deformed structures are known to
dominate the nuclear landscape and symmetry-guided selections are essential. In
addition, the SA-NCSM is suitable for applications to collective states and
rotational bands, including those that lie in high level-density regions of
nuclear spectra. In particular, the SA-NCSM enables one to reach beyond
currently accessible model spaces, which are complete through an $N_{\rm max}$ truncation, 
by reducing such model spaces and augmenting them with configurations beyond $N_{\rm max}$ that are necessary to describe
collective modes.  We have shown that the current implementation of the
SA-NCSM, {\tt LSU3shell}, is highly scalable and efficiently manages the
added computational time associated with tracking symmetries. Moreover, it
exhibits reasonable scaling of the associated memory requirement with $N_{\rm
max}$. These features make SA-NCSM calculations feasible on present massively
parallel computer architectures.
This, in turn, opens the path toward  no-core shell-model descriptions of largely
deformed nuclear states  and cluster substructures as well as heavier nuclei.

\bigskip
\bigskip
%%%%%%%%%%%%%%%%%%%%%%%%%%%%%%%%%%%%%%%%%%%%%%%%%%%%%%%%%%%%%%%%%%%%%%%%%%%%%%%
\centerline{\bf ACKNOWLEDGMENTS}
%%%%%%%%%%%%%%%%%%%%%%%%%%%%%%%%%%%%%%%%%%%%%%%%%%%%%%%%%%%%%%%%%%%%%%%%%%%%%%%
\bigskip

We acknowledge useful discussions with B. R. Barrett and A. M. Shirokov, as well as with D. J. Rowe and C. Bahri.
We are grateful to E. Ng and C. Yang for developments leading to the $J$-scheme statistics for Table \ref{Table:NNZ_12C} and Fig. \ref{Table:NNZ_6Li}.
This work was supported in part by NSF grants OCI-0904874, OCI-0904782, and OCI-0904809.
%NSF0904782,
%% (VERIFY)  -OK? - YES! (JPV)
 %NSF0904874
%% (VERIFIED -- KDL)
%and NSF0500291.
%%  (VERIFY) . -OK?
In addition, this research was supported in part by US DOE Grants  DE-SC0005248, 
% (VERIFIED -- KDL)
DE-FG02-95ER-40934, DESC0008485 (SciDAC/NUCLEI) and DE-FG02-87ER40371, 
%  (VERIFY) . -OK?
by the Czech Science Foundation under Grant No. 16-16772S,
by the Southeastern Universities Research Association, and by the US DOE, Office of
   Advanced Scientific Computing Research,  through the Ames Laboratory,
    operated by Iowa State University under contract No.~DE-AC02-07CH11358.  A portion
of the computational resources were provided by the National Energy Research
Scientific Computing Center, which is supported by the Office of Science of
US DOE under Contract No. DE-AC02-05CH11231, and by an
INCITE award, ``Nuclear Structure and Nuclear Reactions'', from the DOE Office
of Advanced Scientific Computing.  This research also used resources of the Oak
Ridge Leadership Computing Facility at ORNL, which is supported by the DOE
Office of Science under Contract DE-AC05-00OR22725. 
This work is also part of the ``Collaborative Research: Innovative {\it ab initio}
symmetry-adapted no-core shell model for advancing fundamental physics and
astrophysics'' PRAC allocation support by NSF
(award number ACI-1516338), and is part of the Blue Waters
sustained-petascale computing project, which is supported by NSF (awards OCI-0725070 and ACI-1238993) and the state of
Illinois (Blue Waters is a joint effort of the University of Illinois at
Urbana-Champaign and its National Center for Supercomputing Applications).
Portions of this research were conducted with high performance computing
resources provided by Louisiana State University (http://www.hpc.lsu.edu).
%\vfill\eject
%%%%%%%%%%%%%%%%%%%%%%%%%%%%%%%%%%%%%%%%%%%%%%%%%%%%%%%%%%%%%%%%%%%%%%%%%%%%%%%

%%%%%%%%%%%%%%%%%%%%%%%%%%%%%%%%%%%%%%%%%%%%%%%%%%%%%%%%%%%%%%%%%%%%%%%%%%%%%%%
\end{document}